\documentclass[a4paper,fleqn]{cas-dc}

\usepackage[authoryear,longnamesfirst]{natbib}
\usepackage{lineno,hyperref}
\usepackage{ulem}
\usepackage{algorithm2e,setspace}
\usepackage{subfigure}
\usepackage{multirow}
\usepackage{bbding}
\usepackage{amssymb}

\def\tsc#1{\csdef{#1}{\textsc{\lowercase{#1}}\xspace}}
\tsc{WGM}
\tsc{QE}
\tsc{EP}
\tsc{PMS}
\tsc{BEC}
\tsc{DE}

\begin{document}
\let\WriteBookmarks\relax
\def\floatpagepagefraction{1}
\def\textpagefraction{.001}

\shorttitle{Adaptive ship-radiated noise recognition with learnable fine-grained wavelet transform}

\shortauthors{Xie et~al.}

\title [mode = title]{Adaptive ship-radiated noise recognition with learnable fine-grained wavelet transform}                      




%
\author[address1,address2]{Yuan Xie}[
                        style=chinese,
                        orcid=0000-0003-3803-0929]


\ead{xieyuan@hccl.ioa.cn.cn}



\credit{Conceptualization, Methodology, Software, Validation, Formal analysis, Investigation, Data Curation, Writing - Original Draft, Writing - Review \& Editing, Visualization}

\author[address1,address2]{Jiawei Ren}[style=chinese]
\ead{renjiawei@hccl.ioa.cn.cn}
\credit{Methodology, Investigation, Data Curation, Supervision}

\author[address1,address2]{Ji Xu}[style=chinese, orcid=0000-0002-3754-228X]
\ead{xuji@hccl.ioa.cn.cn}
\cormark[1]

\credit{Resources, Writing - Review \& Editing, Supervision, Project administration, Funding acquisition}

\affiliation[address1]{organization={Key Laboratory of Speech Acoustics and Content Understanding, Institute of Acoustics, Chinese Academy of Sciences},
    addressline={No.21, Beisihuan West Road, Haidian District},
    postcode={100190},
    city={Beijing},
    country={China}}
    
\affiliation[address2]{organization={University of Chinese Academy of Sciences},
    addressline={No.80, Zhongguancun East Road, Haidian District}, 
    postcode={100190},
    city={Beijing},
    country={China}}


\cortext[cor1]{Corresponding author}




\begin{abstract}
Analyzing the ocean acoustic environment is a tricky task. Background noise and variable channel transmission environment make it complicated to implement accurate ship-radiated noise recognition. Existing recognition systems are weak in addressing the variable underwater environment, thus leading to disappointing performance in practical application. In order to keep the recognition system robust in various underwater environments, this work proposes an adaptive generalized recognition system - AGNet (\textbf{A}daptive \textbf{G}eneralized \textbf{Net}work). By converting fixed wavelet parameters into fine-grained learnable parameters, AGNet learns the characteristics of underwater sound at different frequencies. Its flexible and fine-grained design is conducive to capturing more background acoustic information (e.g., background noise, underwater transmission channel). To utilize the implicit information in wavelet spectrograms, AGNet adopts the convolutional neural network with parallel convolution attention modules as the classifier. Experiments reveal that our AGNet outperforms all baseline methods on several underwater acoustic datasets, and AGNet could benefit more from transfer learning. Moreover, AGNet shows robust performance against various interference factors.
\end{abstract}




\maketitle

\section{Introduction}
Ship-radiated noise is one of the main contributors to ambient ocean noise~\citep{brooker2016measurement,wang2021noisenet}. Therefore, ship-radiated noise recognition is of great importance in marine acoustics. Recognizing vessels from their radiation noise could be necessary for monitoring maritime traffic and identifying the source of noise in ocean environmental monitoring systems~\citep{fillinger2010towards,sutin2010stevens}. In recent years, the increasing demand has fostered research aiming at building robust ship-radiated noise recognition systems~\citep{li2017denoising,ke2020integrated}.

In previous studies, researchers used acoustic feature extraction to convert signals into non-redundant acoustic features. Das et al.~\citep{das2013marine} applied a cepstrum-based approach to realize marine vessel recognition. Wang et al.~\citep{wang2014robust} used a bark-wavelet analysis combined with the Hilbert-Huang transform. Besides, acoustic features from the audio and speech domains (e.g., Mel frequency cepstrum coefficients) are widely applied to ship-radiated noise recognition tasks and show promising results~\citep{zhang2016feature,khishe2019passive}. However, low-dimensional acoustic features inherently limit the generalization ability, and the recognition performance of classical machine learning models (e.g., support vector machine) is not satisfactory for large-scale data with diverse feature space~\citep{irfan2021deepship}.

With the development of deep learning~\citep{lecun2015deep} and the accumulation of underwater acoustic database~\citep{santos2016shipsear,irfan2021deepship}, recognition algorithm based on deep learning continues to grow in popularity. As reported in the literature, most deep learning methods prefer to use time–frequency-based features for underwater acoustic recognition~\citep{shen2020ship, zheng2021time}. Wang et al.~\citep{wang2018underwater} used LOFAR (Low-Frequency Analysis Recording) to reflect the power spectrum distribution and the change of signals in time and frequency dimensions. Zhang et al.~\citep{zhang2021integrated} applied the short-time Fourier transform (STFT) amplitude spectrum, STFT phase spectrum, and bispectrum features as the input for the convolutional neural network~\citep{krizhevsky2012imagenet}. In addition, spectrograms based on Mel filter banks, Gabor transform, and wavelet transform were applied to the underwater acoustic field to localize the time-frequency domain information~\citep{liu2021underwater,shastri2013time,courmontagne2012time}. Compared with traditional methods, time-frequency-based features contain more comprehensive information, and deep neural networks have high computational efficiency and promising recognition performance~\citep{xie2020time,zhu2021convolutional,liu2021underwater}.

According to the research status, there are several deficiencies in current underwater acoustic recognition systems. First, considering that underwater acoustic signals are complex due to the volatile marine environment, the parameters that cost a lot of time to manually select may be hard to keep optimal in different circumstances, thus causing the poor generalization ability of the recognition system. In addition, for underwater acoustic signals, the information in different frequency bands varies greatly. There is a lot of useless information and interference in the spectrogram. Classifiers in existing work lack the ability to focus attention on valid time-frequency domain information.

In order to solve the above problems existing in recent work, this work proposes an adaptive generalized neural network - AGNet  (\textbf{A}daptive \textbf{G}eneralized \textbf{N}etwork). To enhance the generalization ability of features, AGNet adopts a feature extraction method based on the fine-grained wavelet transform, which could automatically learn wavelet parameters for different center frequencies. This work implements adaptive learning and updating of parameters in a data-driven approach. The parameters of the adaptive generalized network will be influenced by background noise, transmission channel, and other factors so that the recognition model is more suitable for practical application scenarios. Moreover, to focus the attention of the network on valid time-frequency domain information, this work adds parallel convolution attention modules~\citep{chollet2017xception, woo2018cbam} to the conventional neural network, aiming to implement adaptive information mining for the time-frequency spectrograms.

The performance of AGNet on three underwater ship-radiated noise databases is evaluated in this work. AGNet achieves satisfactory recognition accuracy on all datasets: Shipsear (85.48\%), DeepShip (77.09\%), and data collected from Thousand Island Lake (95.76\%). Experiments show that AGNet could maintain excellent performance in the case of low signal-to-noise ratio or low cutoff frequency. Furthermore, AGNet could benefit more from transfer learning due to more learnable parameters. The contributions of our work are summarized as follows:

\begin{figure*}
    \label{fig:framework}
    \centering
    \includegraphics[width=0.7\linewidth]{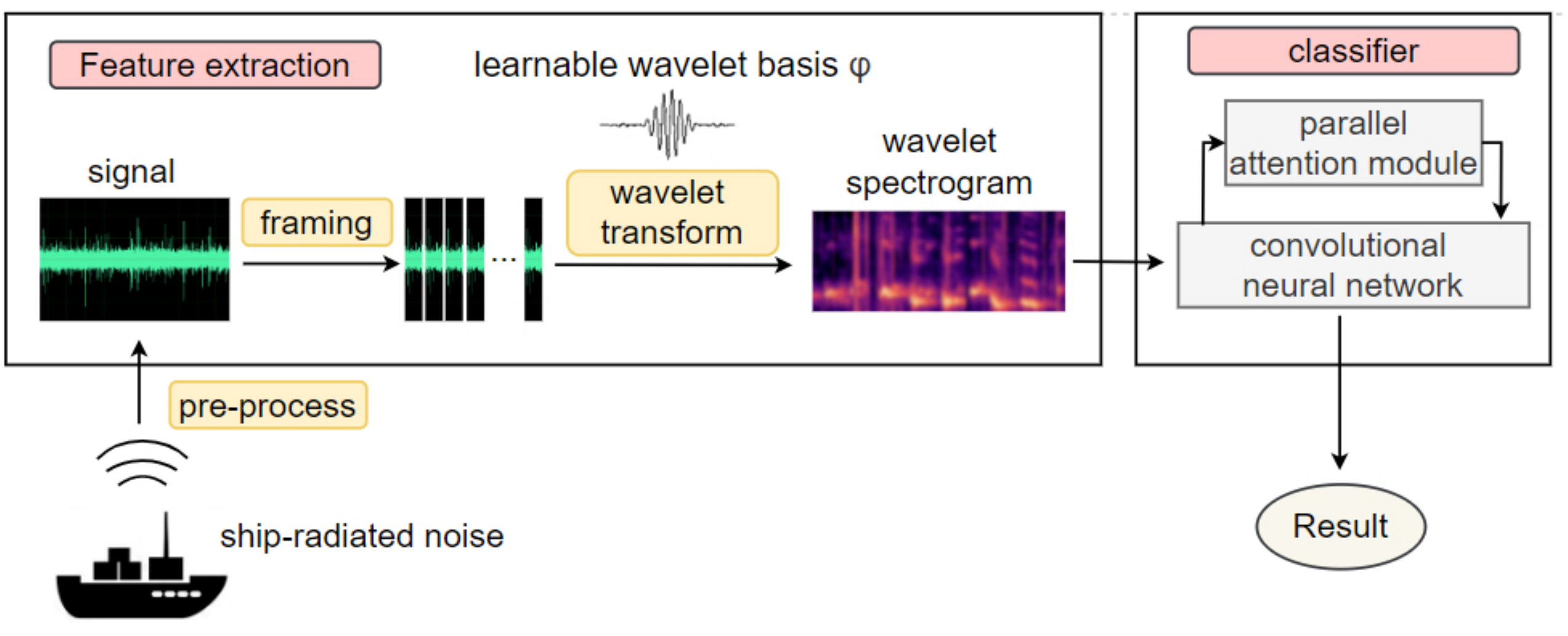}
    \caption{The framework of AGNet. It is an end-to-end system, including pre-processing, wavelet-based feature extraction, and classifier with a parallel attention module. The automatic system receives the ship-radiated noise and outputs the recognition results.}
\end{figure*}

\begin{itemize}
\item AGNet adopts a fine-grained learnable wavelet transform for capturing complex underwater acoustic characteristics. The wavelet parameters of different center frequencies update adaptively in a data-driven manner.

\item To make the classifier focus on the valid information in spectrograms, this work adds parallel convolution attention blocks to the neural network, and adopt the depth-wise separable convolution structure to reduce the parameters.

\item Experiments show that the adaptive generalized network could exhibit satisfactory performance and strong generalization ability on various datasets. More learnable parameters help AGNet benefit more from transfer learning and become robust against high background noise and low cut-off frequency.

\end{itemize}

\section{Methodology}

In this section, the paper gives an overview of the AGNet recognition system and introduces two main innovations of AGNet: learnable fine-grained wavelet transform and parallel convolutional attention module.

\subsection{System overview}

As shown in Figure 1, the process of our system could be divided into three stages: data pre-processing, wavelet-based feature extraction, and attention-based classifier. First, the sonar array will collect the ship-radiated noise. In the data pre-processing stage, mono audio signals could be obtained by array signal processing (e.g., beamforming). After that, audio signals will be framed and transformed into two-dimensional spectrograms by wavelet transform. The classifier will receive the spectrograms and output the classification results. As can be seen from Figure~\ref{fig:framework}, AGNet is an end-to-end system. Parameters of wavelet basis function and classifier will update synchronously during training. The detailed introduction is as follows.

\subsection{Learnable fine-grained wavelet transform}

For underwater acoustic signals, the information in different frequency bands varies greatly. It is necessary to perform differential analysis for frequency bands. In this work, AGNet applies the wavelet transform for adjustable time or frequency resolution. Furthermore, AGNet converts wavelet parameters into learnable ones. It allows the wavelet basis at different center frequencies to learn differential parameters in a data-driven manner, so as to achieve fine-grained wavelet transformation.

\begin{figure*}
    \label{fig:wavelet}
    \centering
    \includegraphics[width=0.7\linewidth]{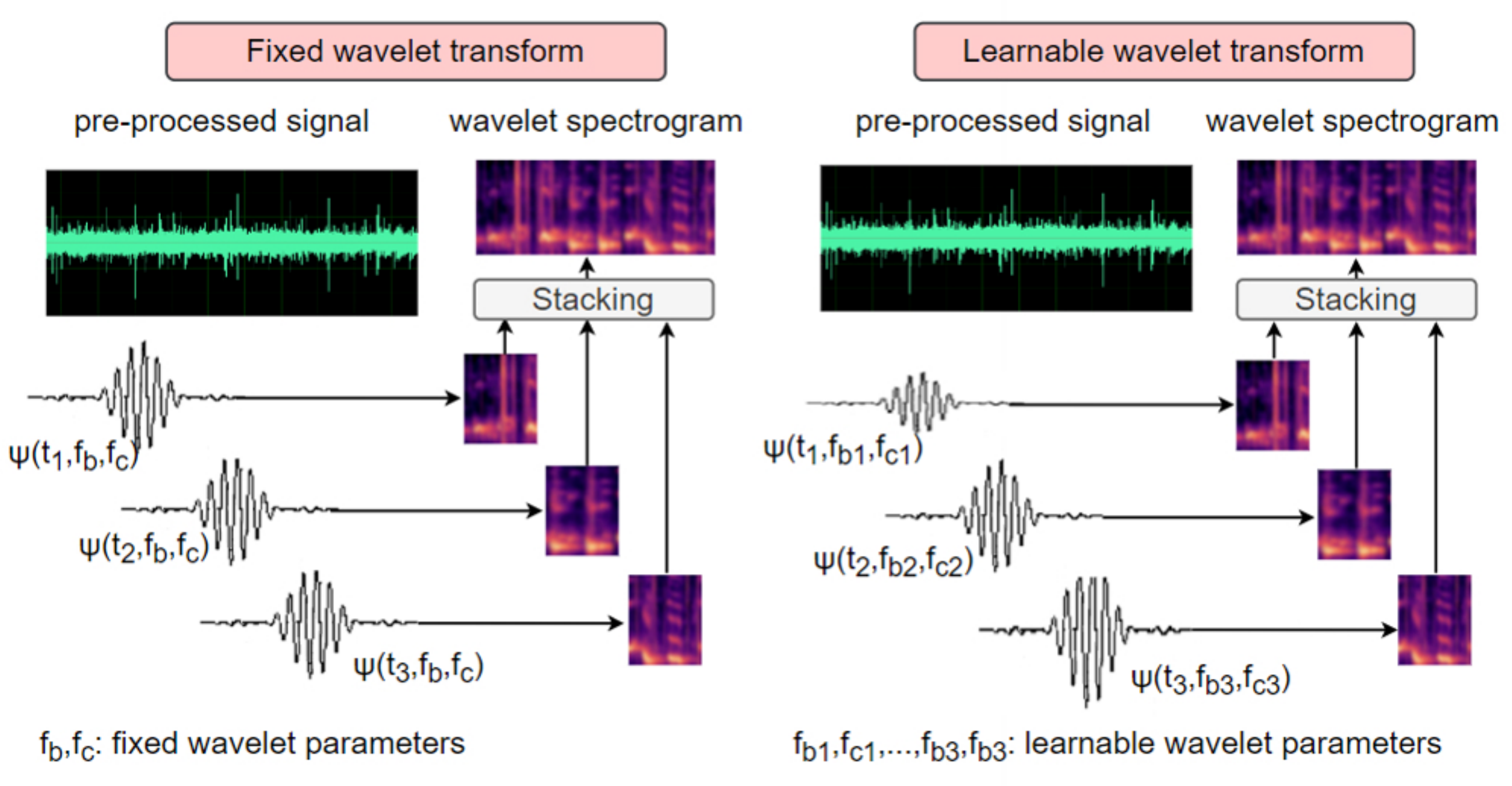}
    \caption{The pipeline of the learnable fine-grained wavelet transform. $t_i$ is the time scale parameter. The figure on the left is the traditional wavelet transform with fixed parameters, and the figure on the right is the learnable wavelet transform adopted by AGNet. AGNet will automatically learn the optimal $f_b$, $f_c$.}
\end{figure*}

\begin{equation}
    X(n,\tau) = \sum_{-\infty} ^ {+\infty} x[n] w[n-\tau]\psi(n)
\end{equation}

Denote the input signal as $x$, the point of the discrete sampling sequence as $n$, and the window function as $w[\cdot]$. The commonly used wavelet transform formula is shown in Equation 1. $\tau$ is the time center parameter, and $\psi(\cdot)$ is the wavelet basis function.

It is necessary to choose an appropriate wavelet basis function according to the signal characteristics. Underwater acoustic signals tend to be characterized by high background noise and low-frequency. According to the survey, this work choose three wavelet basis functions that are widely applied in the underwater acoustic field~\citep{wang2013time,chen2019underwater,kalpana2014study,patil2014wavelet}. Besides, they are also shown to be effective on low-frequency acoustic tasks~\citep{wu2019deep,bhavaraju2010comparative,guzhov2021esresne}. Denote the bandwidth as $f_b$, the center frequency as $f_c$, and the order as $m$. The wavelet basis function could be formulated as follows:

\begin{itemize}
\item Complex Morlet wavelet (Cmor):
Cmor is a commonly used wavelet basis in practical applications~\citep{wang2013time,chen2019underwater}.In addition, it is also used in tasks that require high resolution for low-frequency bands, such as atrial fibrillation detection. The basis function is described as follows:
\begin{equation}
    \psi(n) = \frac{1}{\sqrt{\pi \cdot f_b }}exp(\frac{- n^{2}}{f_b})exp(2\pi i f_c  n)
\end{equation}

\item Complex Shannon wavelet (Shan):
Complex Shannon wavelet is applied to de-noise underwater acoustic communication~\citep{kalpana2014study}. It has a good effect on noise suppression in the complex marine environment. The basis function is described as follows:
\begin{equation}
    \psi(n) = \sqrt{f_b}(sinc(f_b n)) \exp(2i\pi f_c n).
\end{equation}

\item Complex frequency B-spline wavelet (Fbsp):
Similar to Shan wavelet, Fbsp wavelet also exhibits promising performance on de-noising the noisy underwater environment~\citep{patil2014wavelet}. The basis function is described as follows:
\begin{equation}
    \psi(n) = \sqrt{f_b}(sinc(\frac{f_b n}{m}))^m \exp(2i\pi f_c n)
\end{equation}

\end{itemize}

The adjustable parameters of the wavelet function include order parameter $m$, bandwidth parameter $f_b$ and wavelet center frequency $f_c$. Normally, $m$ and $f_b$ are set firmly. In the marine environment, signals are highly influenced by background noise, transmission channels, and other factors. Some frequency bands contain valid information, while others suffer from interference such as noise. The wavelet basis with fixed parameters seems to be sub-optimal.

To pursue better generalization ability, this work utilizes the end-to-end network to update wavelet parameters, thus realizing the data-driven feature extraction. AGNet enables $m$, $f_b$, and $f_c$ to participate in the gradient calculation and back propagation. As depicted in Figure 2, AGNet could adaptively learn wavelet parameters $m$, $f_b$ for different center frequencies $f_c$.


\subsection{Attention-based classifier}

\begin{figure*}
    \centering
    \scalebox{1}{
    \subfigure[The structure of attention-based ResNet. All attention blocks are parallel to the bottleneck layers. ``conv'' represents the convolution layer. The classifier is fed with the wavelet spectrogram and outputs the recognition results.]{
        \begin{minipage}[b]{1\textwidth}
        \includegraphics[width=0.75\linewidth]{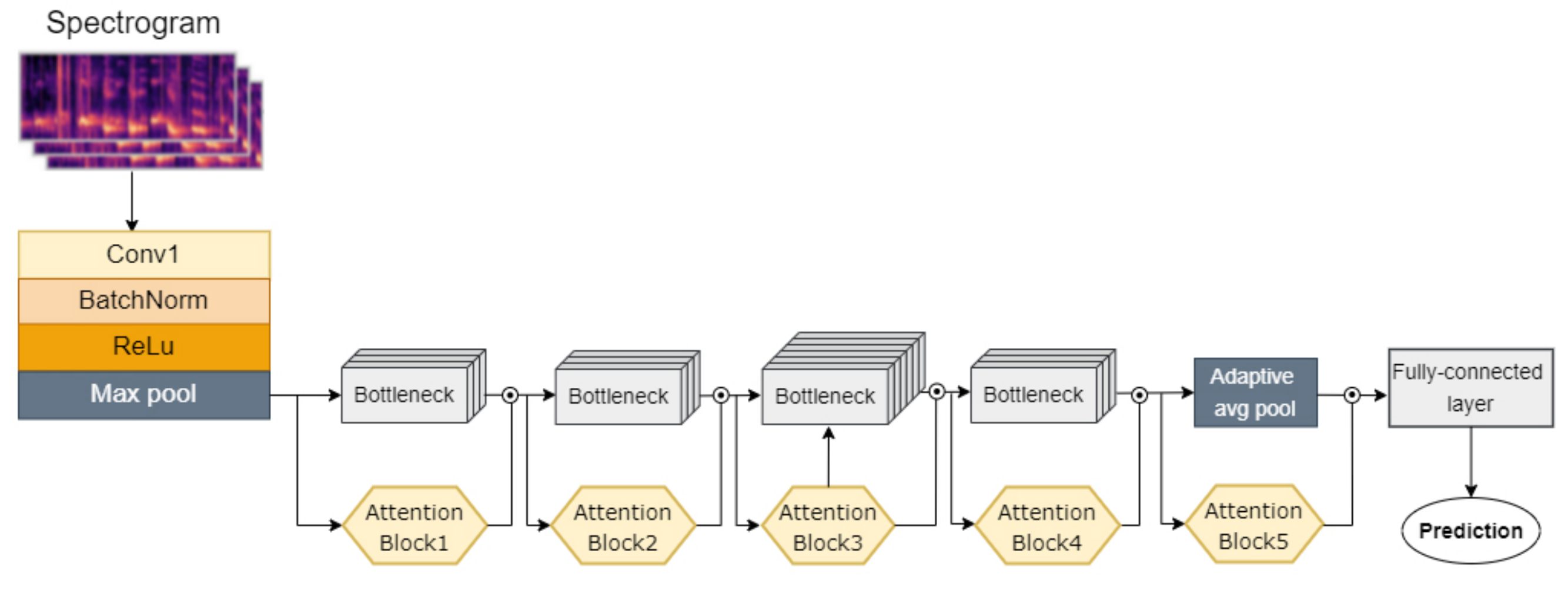}
        \centering
        \end{minipage}}
    }
        
    \scalebox{1}{
    \subfigure[The structure of attention blocks (left) and bottleneck layers (right). ``conv'' represents the convolution layer, while ``1x1,3x3''represents the size of the convolution kernel.]{
        \begin{minipage}[b]{1\textwidth}
        \includegraphics[width=0.7\linewidth]{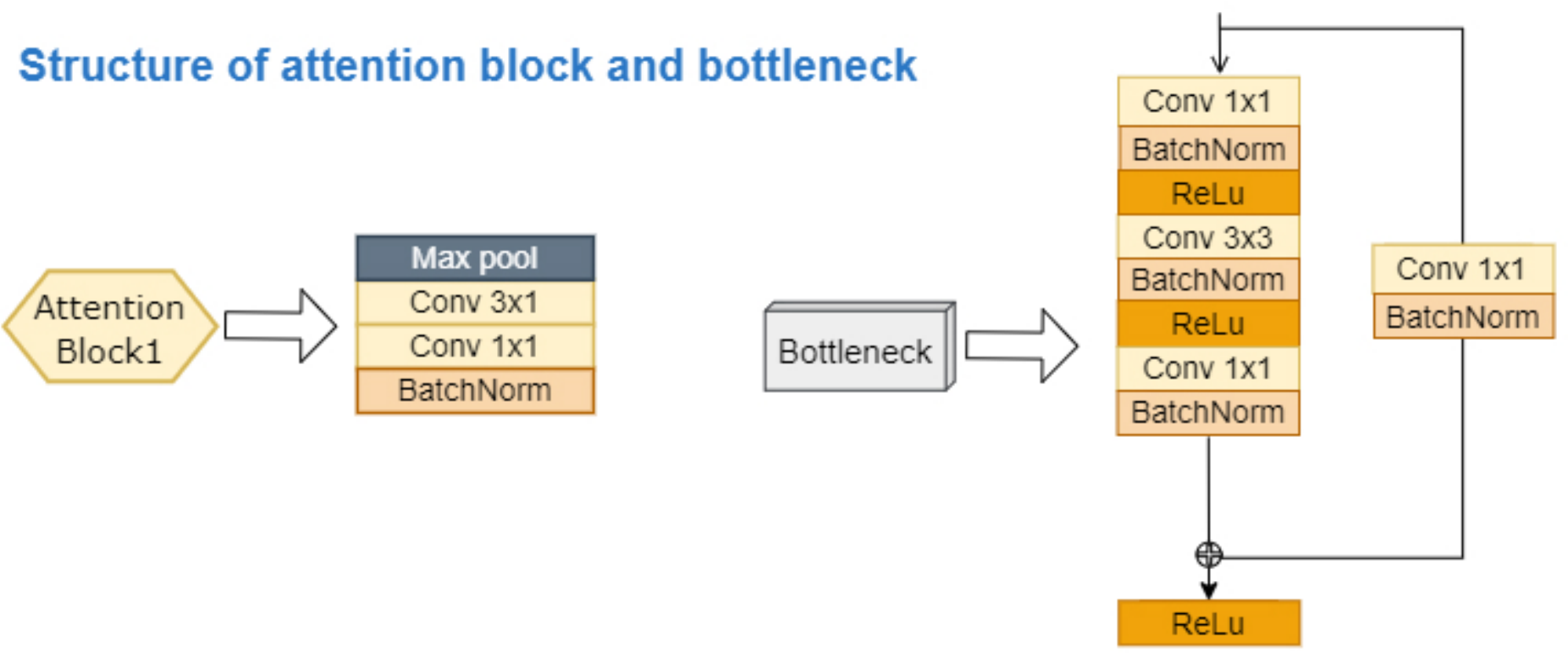}
        \centering
        \end{minipage}}
        }

    \caption{The structure of the attention-based classifier. Figure (a) is the overall structure of the classifier, while figure (b) is the detailed structure of the attention block and the bottleneck layer in ResNet.}
    \label{fig:classifier}
\end{figure*}

As a widely used convolutional neural network, ResNet~\citep{he2016deep} has been a common baseline for classification networks since its inception. For AGNet, this work follows the ResNet backbone and make some optimization for the particularity of the underwater signal spectrograms.
As illustrated in Figure \ref{fig:classifier}(a), our classifier consists of a convolution layer stacked with a max-pooling layer, followed by four residual layers, an average pooling layer, and a fully-connected layer. Following the structure in ResNet, each residual layer contains the stack of the bottleneck layers that include convolution layers, batch normalization layers, and skip-connection. Based on the ResNet backbone, AGNet adds several parallel convolution attention blocks.

The structure of the attention block is shown in Figure \ref{fig:classifier}(b), each attention block includes the max-pooling operation, followed by a depth-wise separable convolution~\citep{chollet2017xception} stacked with batch normalization. The depth-wise separable convolution decouples the conventional convolution. Each filter of the depth-wise convolution layer (conv 3x1) is convolved with only one channel of the input, and then the point-wise convolution layer (conv 1x1) merges the results of the previous convolution layer by the 1x1 convolution kernel. The structure can greatly reduce the number of parameters and computation time~\citep{howard2017mobilenets}.

Each attention block corresponds to a residual layer. As shown in Figure \ref{fig:classifier}(a), the attention block receives the same input as the corresponding residual layer. The residual layer outputs feature maps, while the attention layer outputs the attention matrix. The attention matrix does not change the dimension of feature maps, it only changes the weights of feature maps by multiplying. The attention block could help the network focus the attention on the informative parts (e.g., beat or rhythm).

\subsection{Training process}
As an end-to-end system, AGNet accepts raw audio sequences directly as input. The wavelet transformation layer will convert audio sequences into two-dimensional wavelet spectrograms. Then, spectrograms will be fed into the attention-based classifier. AGNet calculates the loss between the predicted value and the ground truth (label annotations) by the cross-entropy loss function. The parameters of all modules (wavelet basis funtion, ResNet-based network, attention blocks) will synchronously update according to the gradients.

When handling tasks with different classes, it only needs to modify the fully-connected layer at the end of the classifier according to the specific task. Other structures and training processes remain unchanged.

\section{Experiment Setup}
\subsection{Datasets}

1. Shipsear~\citep{santos2016shipsear} is an open-source database of underwater recordings of ship and boat sounds. The database is currently composed of 90 records representing sounds from 11 vessel types. It consists of nearly 3 hours of recordings. Considering that too little data is difficult to cut into the form of ``train, validation, test'',this work selects a subset of 9 categories (Dredger, Fishboat, Motorboat, Musselboat, Naturalnoise, Oceanliner, Passengers, Roll on/Roll off Ship, Sailboat) in Shipsear for the recognition task. Currently, most of the work on Shipsear divides all types into five classes by size. The mapping relation is shown in Table.\ref{tab:shipsear}. Although AGNet still applies vessel types as the input during training, the final accuracy according to the official classification criteria (Class A - Class E) will be calculated to facilitate comparisons with other work on Shipsear~\citep{santos2016shipsear}. 

\begin{table}[ht]
    \centering
	\caption{\label{tab:shipsear} Following the operations in~\citep{santos2016shipsear}, the 9 vessel types (Dredger, Fishboat, Motorboat, Musselboat, Naturalnoise, Oceanliner, Passengers, Roll on/Roll off Ship, Sailboat) were merged into 4 experimental classes (based on vessel size) and 1 background noise class.}
		\scalebox{1}{\begin{tabular}{cc}
		    \hline
		    Class  & Types\\
		    \hline
			A & Fishboat, Musselboat, Dredger\\
			B & Motorboat, Sailboat\\
			C & Passengers\\
			D & Oceanliner, Roll on/Roll off Ship\\
			E & Naturalnoise\\
			
			\hline
		\end{tabular}}
\end{table}

2. DeepShip~\citep{irfan2021deepship} is an open source underwater acoustic benchmark dataset, which consists of 47 hours and 4 minutes of real-world underwater recordings of 265 different ships belonging to four classes (cargo, passenger ship, tanker, and tug).

3. Data collected from Thousand Island Lake (DTIL)~\citep{ren2019feature} is a dataset collected from Thousand Island Lake, which contains multiple sources of interference. The two types of targets are speedboat and experimental vessel. The dataset contains 330 minutes of the speedboat and 285 minutes of the experimental vessel. 

In addition, this work also uses the dataset - AudioSet~\citep{gemmeke2017audio} in the audio field for transfer learning. AudioSet is a large-scale sound dataset that provides approximately 1.8 M sound clips organized into 527 classes in a multi-label manner. Among them, a small amount of data contains several vessel records (e.g., motorboat, fishboat ,etc.).

\subsection{Unification and division of training data}

The three datasets (Shipsear, DeepShip, DTIL)used in this experiment have different sampling rates (52734Hz, 32000Hz, 17067Hz). The audio files of the three datasets are downsampled to 16000Hz. Each full audio is cut into 30-second segments and adjacent segments have an overlap of 15 seconds. Different segments of the same audio sequence will be assigned to the same fold to ensure there is no overlap between training and test sets.

It is necessary to ensure that the data in test sets is irrelevant enough to the data in training sets. Only then can the reported accuracy reflect the recognition ability and generalization performance. Due to the lack of relevant information, it is hard to manually give a reasonable enough split for Shipsear and DeepShip. So this work uses 4-fold cross-validation to ensure that our results are credible and persuasive. The final recognition accuracy is the average of the four folds. For DTIL, authors have recorded relevant information when collecting data, so a reliable train/test split could be given. All test data are collected individually over a certain period.

\subsection{Parameters setup}

For all framing operations in experiments (including AGNet and all baseline methods), this work sets the frame length as 100ms and the frame shift as 50ms. When implementing the baseline of Mel-filter banks, the number of filter banks is set to 300. For wavelet parameters, this work follows the setup in ~\citep{guzhov2021esresne} and set $m=0, f_b=1$$\footnote{If assigning $m=0$ and $f_b=$1 for Shan wavelet or Fbsp wavelet, the wavelet basis funtion $\psi(n) = \exp(2i\pi f_c n)$ is identical to the
inverse DFT. Here, the basis function is the same as that of STFT, which can be regarded as a good enough state.}$. For learnable wavelet parameters, we also apply them as initial values.

During training, AGNet use the Adam optimizer~\citep{kingma2014adam} with weight decay regularization. The learning rate is set to 5e-4 and the weight decay is set to 1e-6 for all experiments. All models are trained for 100 epochs on 4 V100 GPUs.

\section{Results and analysis}

In this work, we first demonstrate the practicability of wavelet-based front end with learnable fine-grained parameters. Then, ablation experiments prove that the attention blocks of the classifier module could help improve the recognition accuracy. Furthermore, we find that AGNet could benefit a lot from transfer learning. Finally, we evaluate the robustness of AGNet, showing that it is less affected by colored noise and low cutoff frequency than other recognition systems. The specific results and detailed analysis are presented in the following sub-sections.

All experiments are implemented on Shipsear (5-class classification), DeepShip (4-class classification: cargo, passenger ship, tanker, tug), and DTIL (2-class classification: speedboat, experimental vessel).

\subsection{Learnable wavelet transform}

As an indispensable part of the system, the feature extraction front-end has an appreciable impact on the generalization ability. In this work, we experimentally evaluate learnable fine-grained wavelets. To build weak baselines, we implement low-dimensional acoustic features and time-frequency-based features with manually set parameters. Additionally, we take the wavelet spectrogram with fixed parameters as strong baselines. For a fair comparison, other modules in the recognition system keep consistent except for the feature extraction front-end.

\begin{table}[ht]

	\caption{\label{tab:table1} Experiments on the recognition system with various feature extraction front-ends. All of them use ResNet50 as the classifier uniformly. Thereinto, time-domain features include energy, zero-crossing wavelength, zero-crossing wavelength difference, peak-to-peak amplitude, etc. We choose accuracy as the metric, and all results are measured in 30 seconds rather than the full audio clip. ``spec'' in the table means ``spectrogram''.}
	\centering
	\scalebox{0.81}{
		\begin{tabular}{lccc}
            \hline
			Features & Shipsear & DeepShip & DTIL\\
			\hline
			
			Time-domain features  &63.20& 54.83&79.84 \\
			Demon~\citep{lu2020fundamental} & 70.23& 53.11& 84.46\\
			Mel filter banks~\citep{zhang2016feature} &76.23 & 61.45& 92.48\\
			Vggish~\citep{gemmeke2017audio} & 75.60 & 61.77 & 90.40\\
			STFT spec~\citep{zhang2021integrated} & 76.23 & 57.21& 92.66\\
			Gabor spec~\citep{shastri2013time}&76.49& 60.53 & 90.96\\
			\hline
			Wavelet spec (fixed Cmor $f_b=1$) & 76.79 & 64.99 & 92.13\\
			Wavelet spec (fixed Shan $f_b=1$) & 76.58 & 65.75 & 92.01\\ 
			Wavelet spec (fixed Fbsp $m=0, f_b=1$) & 76.58 & 66.02 & 92.79\\
			\hline
			Wavelet spec (learnable Cmor) & 77.07 & 66.67 & 93.07\\
			Wavelet spec (learnable Shan) & \textbf{77.76}  & 67.46 & 93.19\\
			Wavelet spec (learnable Fbsp) &77.38 &\textbf{67.92} & \textbf{93.23}\\
            \hline

		\end{tabular}}
\end{table}

As shown in Table \ref{tab:table1}, the recognition performance of low-dimensional acoustic features and time-frequency features is not robust enough. For features that perform relatively well, Mel filter banks could only achieve sub-optimal performance on three databases. Meanwhile, the accuracy of systems based on time-frequency-based transform (STFT spectrogram, Gabor spectrogram) is far from satisfactory on DeepShip. According to our analysis, for different sources of underwater acoustic data, hand-crafted parameters (e.g., number of filter banks, FFT numbers, frame length) make it difficult for models to keep a competitive performance. Well-chosen parameters may tend to be inappropriate when the marine environment changes.

Then, it could be seen from the table that the wavelet transform spectrogram is obviously better than the STFT and Gabor spectrograms on three databases. It proves that the adaptive resolution of wavelet transform in the frequency domain is beneficial for the recognition task. Moreover, learnable wavelet spectrograms show better and more robust performance than fixed wavelet spectrograms. Experiment shows that the flexible learning of fine-grained wavelet parameters plays a positive role in the recognition system.

Besides, this work also explores the influence of different wavelet basis functions on the results. Overall, the performance of the complex frequency B-spline wavelet (Fbsp) is slightly better than others. The addition of more learnable parameters (order $M$) makes it possible to draw more informative spectrograms with the help of training data when processing complex underwater acoustic signals. So in all subsequent experiments, the Fbsp wavelet is taken as our wavelet basis uniformly.

\subsection{Visualization and analysis of fine-grained parameters}

For a detailed analysis of fine-grained wavelet parameters, the learnable parameters $m, fb_b$ are visualized in Figures 4 - 5. Among them, the learnable parameter $f_c$ increases almost linearly with the increase of the time center parameter of the wavelet transform. For ease of analysis, we take the value of $f_c$ as the horizontal axis.

\begin{figure*}
    \centering
    \subfigure[Data analysis of $m$ on three datasets.]{
        \begin{minipage}[b]{1\textwidth}
        \includegraphics[width=0.75\linewidth]{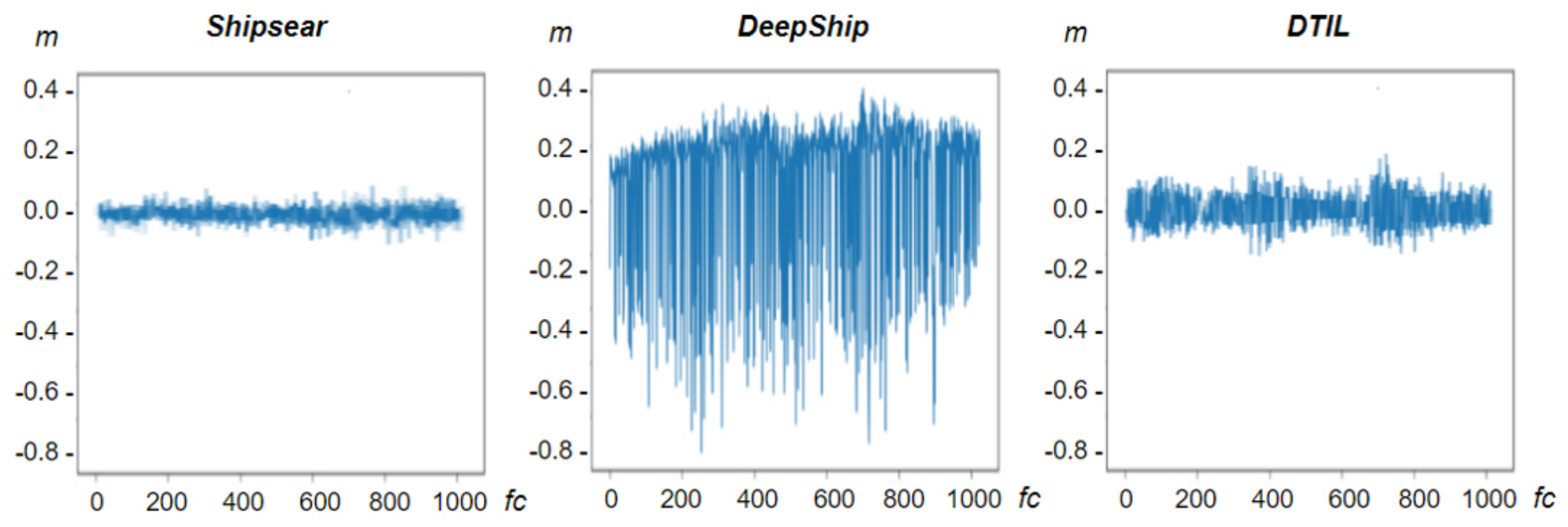}
        \centering
        \end{minipage}}

    \subfigure[Data analysis of $f_b$ on three datasets.]{
        \begin{minipage}[b]{1\textwidth}
        \includegraphics[width=0.75\linewidth]{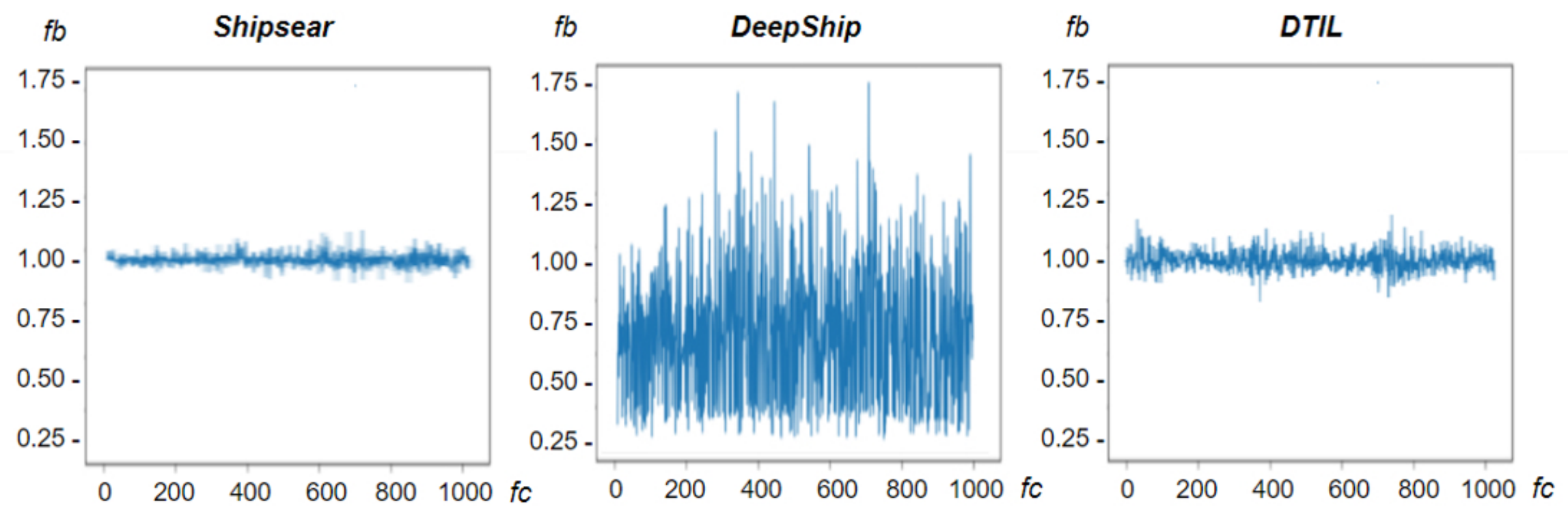}
        \centering
        \end{minipage}}
    \caption{Visualization of learnable parameters. We take the value of $f_c$ as the horizontal axis and the value of $m, f_b$ as the vertical axis.}
    \label{fig:para1}
\end{figure*}

As depicted in Figure 4, it shows the learned values of $m, f_b$ on three datasets. The visualization results reveal that AGNet tends to learn significantly different parameter values on the three datasets. Figure 4 shows that $m$ and $f_b$ do not fluctuate significantly as $f_c$ changes on Shipsear and DTIL ($m$ is approximately 0 and $f_b$ is approximately 1). At this point, the wavelet basis function $\psi(n) = \exp(2i\pi f_c n)$ is similar to the inverse DFT. Previous work has proven it a good choice for clean audio signals~\citep{guzhov2021esresne}(e.g., Shipsear and DTIL). However, on DeepShip with complex background noise, $m$ and $f_b$ fluctuate wildly. The parameter values of adjacent frequency points tend to have a large gap. At this point, the advantages of learnable fine-grained wavelet parameters come to the fore. Taking the Fbsp wavelet basis function as an example, compared to fixed parameters, the learnable parameters on DeepShip could improve the accuracy by 1.9\%, while on Shipsear and DTIL, the accuracy can only get 0.8\% and 0.44\% accuracy benefits respectively.

According to our analysis, considering the complexity of the marine environment, sonar systems, and transmission channels, the underwater acoustic signals from different sources usually vary greatly. Thus for signals from different sources, the values of wavelet parameters vary greatly. Besides, the underwater acoustic signals are periodic, despite numerous interference factors. As can be seen in Figure 4, the values of $f_b$ will only fluctuate within a certain range. The parameters obtained by adaptive learning show that the AGNet is not inclined to set different bandwidths at low frequency and high frequency to pursue different frequency domain resolutions as traditional wavelet transforms. In addition, the wavelet parameters whose numerical scale is limited to a certain range will not affect the robustness of the model. It makes anomalies (e.g. unexpected noise sources, Pulse signals, etc.) have less impact on the model.




\begin{figure*}
    \centering
    \subfigure[Data analysis of $m$ on Shipsear.]{
        \begin{minipage}[b]{1\textwidth}
        \includegraphics[width=0.75\linewidth]{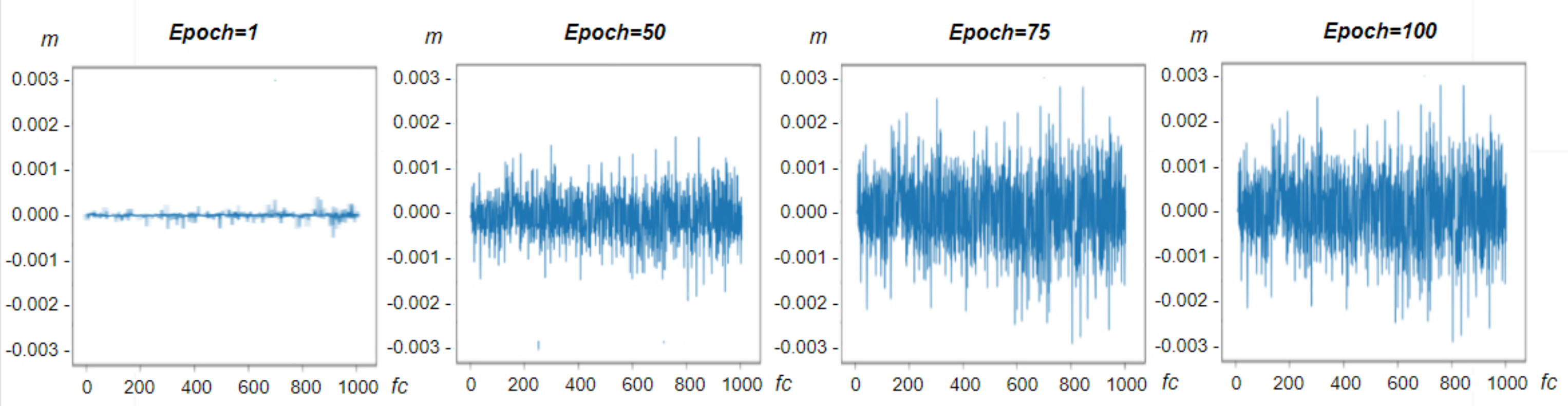}
        \centering
        \end{minipage}}

    \subfigure[Data analysis of $f_b$ on Shipsear.]{
        \begin{minipage}[b]{1\textwidth}
        \includegraphics[width=0.75\linewidth]{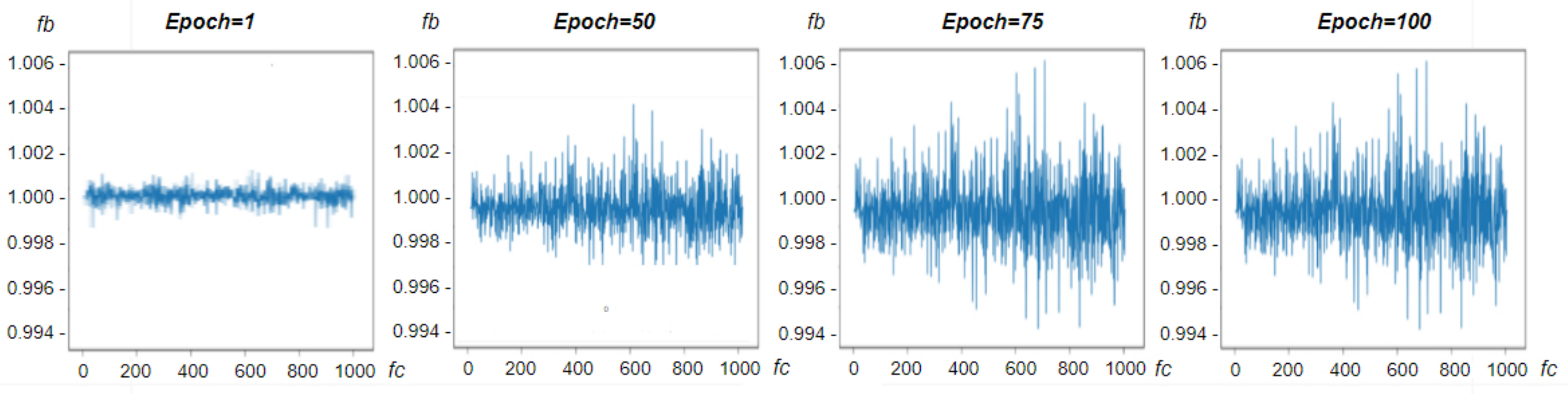}
        \centering
        \end{minipage}}
    \caption{The learning process of $m$ and $f_b$ on Shipsear. The parameters are visualized after epoch1, epoch50, epoch75 and epoch100.}
    \label{fig:para3}
\end{figure*}

After that, the learning process of parameters is visualized. Figure 5 shows that the parameters $m$ and $f_b$ are not updated significantly in the initial stage of training (epoch=1) due to the appropriate initialization. As the training stage goes on, AGNet gradually learns the optimal time-frequency domain resolution assignment from training data. For Shipsear, the wavelet parameter learning of AGNet will reach convergence around epoch 70$\sim$80.

\subsection{Role of attention blocks}

\begin{table}[ht]
	\caption{\label{tab:table2} Ablation study of the attention module. For features, We choose Mel filer banks as the weak baseline, and fixed wavelet (fbsp) as the strong baseline. The backbone of the classifier is ResNet50. We add attention blocks to ResNet50 to conduct the ablation experiments.}
	\centering
	\scalebox{0.73}{
		\begin{tabular}{cccccc}
            \hline
			Features& Classifier & Shipsear& DeepShip & DTIL\\
			\hline

			Mel filter banks &ResNet &76.23& 61.45 & \textbf{92.48}\\
            Mel filter banks &ResNet+Attention &\textbf{76.43}& \textbf{63.25} & 92.47\\
            
			Fixed wavelet (Fbsp)&ResNet & 76.58 & 66.02 & 92.79\\
            Fixed wavelet (Fbsp)&ResNet+Attention & \textbf{79.49} & \textbf{68.03} & \textbf{93.51}\\
            
			Learnable wavelet (Fbsp) &ResNet&77.38 &67.92 & 93.23\\
			Learnable wavelet (Fbsp) &ResNet+Attention&\textbf{79.65} &\textbf{70.47}&\textbf{94.48}\\
            \hline

		\end{tabular}}
\end{table}

An ablation experiment is conducted on the parallel convolutional attention blocks and evaluate the performance on three datasets. Three features are selected for comparison according to the results in Table 2. They are Mel-filter banks with the best recognition accuracy among low dimensional features, fixed fbsp wavelet with the best accuracy among time-frequency-based features, and learnable fbsp wavelet used by AGNet. Table 3 reveals that the attention module could improve the recognition performance of AGNet. For wavelet spectrogram with learnable parameters, the recognition accuracy could improve by 2.27\% on Shipsear and 2.55\% on DeepShip.

 It is worth noting that time-frequency features could benefit more from attention blocks than low-dimensional features. Low-dimensional features compress the dimension in pursuit of more valid information. So adding attention modules to such low-dimensional and informative features will not bring about significant improvement. Fixed or learnable time-frequency features tend to contain more comprehensive information, but a large part of the spectrograms are invalid information or interference. At this point, the superiority of the attention module begins to emerge.


\subsection{Experiments on transfer learning}

\begin{table}[ht]
	\caption{\label{tab:table3} Experiments on transfer learning. All classifiers uniformly use ResNet50 with convolution attention blocks. $\times$ means training from scratch, while $\checkmark$ means pre-training on AudioSet.}
	\centering
	\scalebox{0.75}{
		\begin{tabular}{ccccc}
            \hline
			Features& Transfer weights  & Shipsear & DeepShip & DTIL\\
			\hline
            Mel Filter Banks &$\times$ &76.43& 63.25& 92.47\\
             &\checkmark & \textbf{79.01} & \textbf{66.90}& \textbf{93.07}\\

            Fixed wavelet (Fbsp)&$\times$  & 79.49 & 68.03 & 93.51\\
            &\checkmark&\textbf{84.03} & \textbf{74.26} & \textbf{95.23}\\

			Learnable wavelet (Fbsp) &$\times$& 79.65&70.47&94.48\\
			&\checkmark&\textbf{85.48}& \textbf{77.09} & \textbf{95.76}\\
            \hline

		\end{tabular}}
\end{table}

\begin{figure*}
    \centering
    \subfigure[Confusion matrix on Shipsear.]{
        \begin{minipage}[b]{1\textwidth}
        \includegraphics[width=0.75\linewidth]{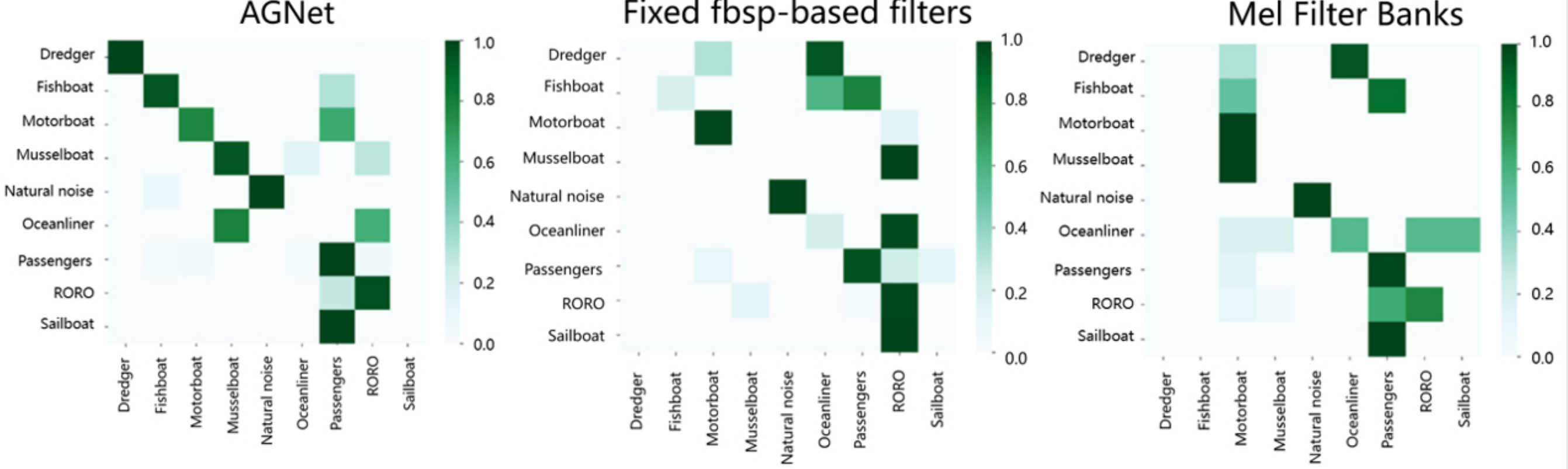}
        \centering
        \end{minipage}}

    \subfigure[Confusion matrix on Deepship.]{
        \begin{minipage}[b]{1\textwidth}
        \includegraphics[width=0.75\linewidth]{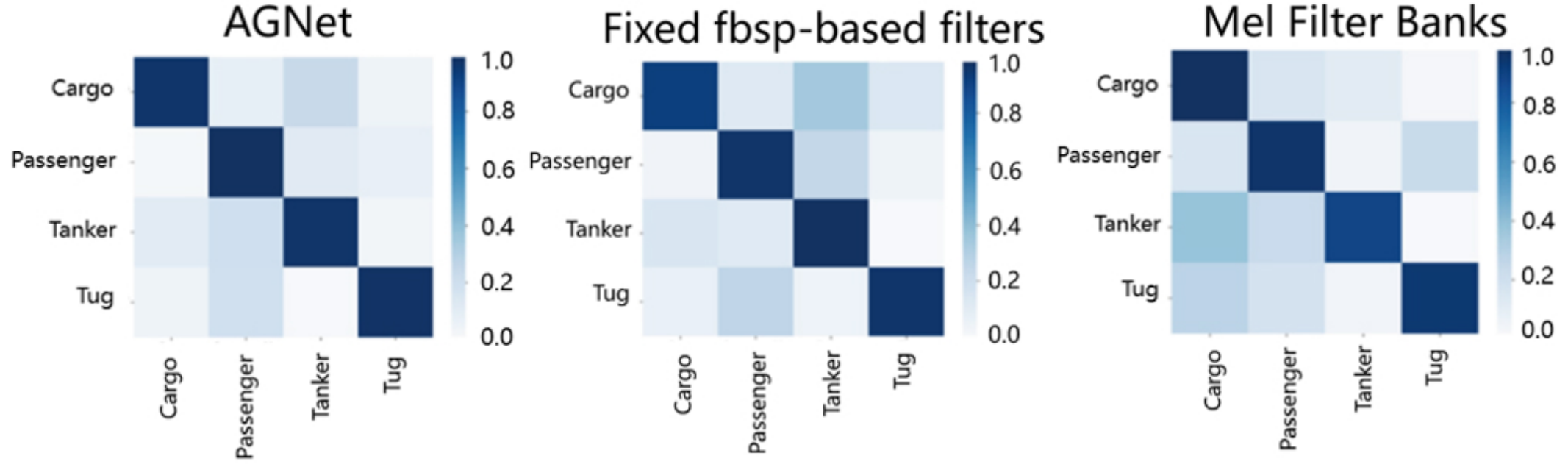}
        \centering
        \end{minipage}}
    \caption{The confusion matrices of the best model of Mel filter banks, fixed wavelet (fbsp) and learnable wavelet (fbsp). The color depth indicates the number of test samples recognized as the corresponding category. Samples on the diagonal are correctly recognized.}
    \label{fig:confusion}
\end{figure*}

Based on a large-scale dataset - AudioSet, we conduct experiments on transfer learning. With the help of AudioSet, we pre-train the AGNet on the source domain (audio domain) and then transfer the model weight as the initial value on the underwater acoustic recognition task. To be more specific, we first set the last fully-connected layer of the classifier as audio task-related (according to AudioSet, the output dimension is 527), then pre-train the model on AudioSet. After training for 100 epochs, AGNet reaches convergence and the weight will be stored in checkpoints. When dealing with underwater acoustic recognition tasks, AGNet will load the pre-trained weights to achieve knowledge transfer. All modules will get initialized with transfer weights except the fully connected layer which needs to be modified. The vast amount of knowledge in AudioSet could be used as prior information to help AGNet to converge faster and better.

As depicted in Table 4, transfer learning improves the performance of AGNet significantly. Compared with baseline methods, the fbsp-based wavelet with more learnable parameters could benefit more from transfer learning. In particular, on DeepShip and Shipsear, transfer learning could lead to the gain of approximately 6\% and 6.5\% respectively, which is quite promising. According to our analysis, there exist some commonalities between the source domain (audio) and the target domain (underwater acoustic). Therefore, parameters of AGNet could get better initialized by rich prior knowledge. In terms of machine learning, it could make the distribution of various types in the embedding space more reasonable. It could make up for the problem of scarce data in the target domain, thus alleviating the issue that models are easy to fall into local optimum.

At this point, the best results for AGNet and the two baseline methods on three databases are achieved. To intuitively demonstrate AGNet's satisfactory recognition performance, the confusion matrices of AGNet and the two baseline methods are plotted on Shipsear and DeepShip. As depicted in Figure 6, the recognition performance of AGNet is significantly ahead of baseline methods. It shows the satisfactory recognition ability on categories that are difficult for baseline methods to recognize (e.g., Dredger, Fishboat, Musselboat on Shipsear).

\subsection{Robustness test}
In ocean environments and transmission channels, various interference factors may cause poor signal quality. High background noise and low cut-off frequency are inevitable. For further study, the robustness of AGNet is tested to evaluate its value in practical applications. Simulated interference are added to all training data and test data. Experimental results are presented in the following sub-sections.

\subsubsection{Test on low Signal to Noise Ratio (SNR)}

First, we implement experiments by adding noise to the audio signals. Inspired by related research on ship-radiated noise~\citep{hazelwood2005estimation}, this work uses colored noise (e.g., red noise) to simulate sea ambient noise instead of traditional Gaussian white noise. Three methods are evaluated on Shipsear and DeepShip with different signal-to-noise ratios. Figure~\ref{fig:robust}(a) shows that the fbsp wavelet could improve the model’s robustness against additive colored noise. From Figure~\ref{fig:robust}(a), when the SNR decreases, the recognition accuracy of AGNet degrades relatively slowly, which is better than other baseline methods. Without adding any noise, AGNet has 6.47\% and 10.19\% performance advantages on Shipsear and DeepShip over the  Mel filter banks baseline. When the noise gradually increases until the signal-to-noise ratio is 0, the gap widens to 19.31\% and 20.48\%. It could be attributed to the good time-frequency localization characteristics of the wavelet transform, which has a positive effect on noise suppression. When the signal-to-noise ratio continues to decrease, the gap will not widen because the signal quality is insufficient for systems to make accurate recognition.


\subsubsection{Test on low cut-off frequency}

\begin{figure*}
    \centering
    \scalebox{1}{
    \subfigure[Test on low Signal to Noise Ratio. On the left are the results on Shipsear and the right are the results on DeepShip.]{
        \begin{minipage}[b]{1\textwidth}
        \includegraphics[width=0.75\linewidth]{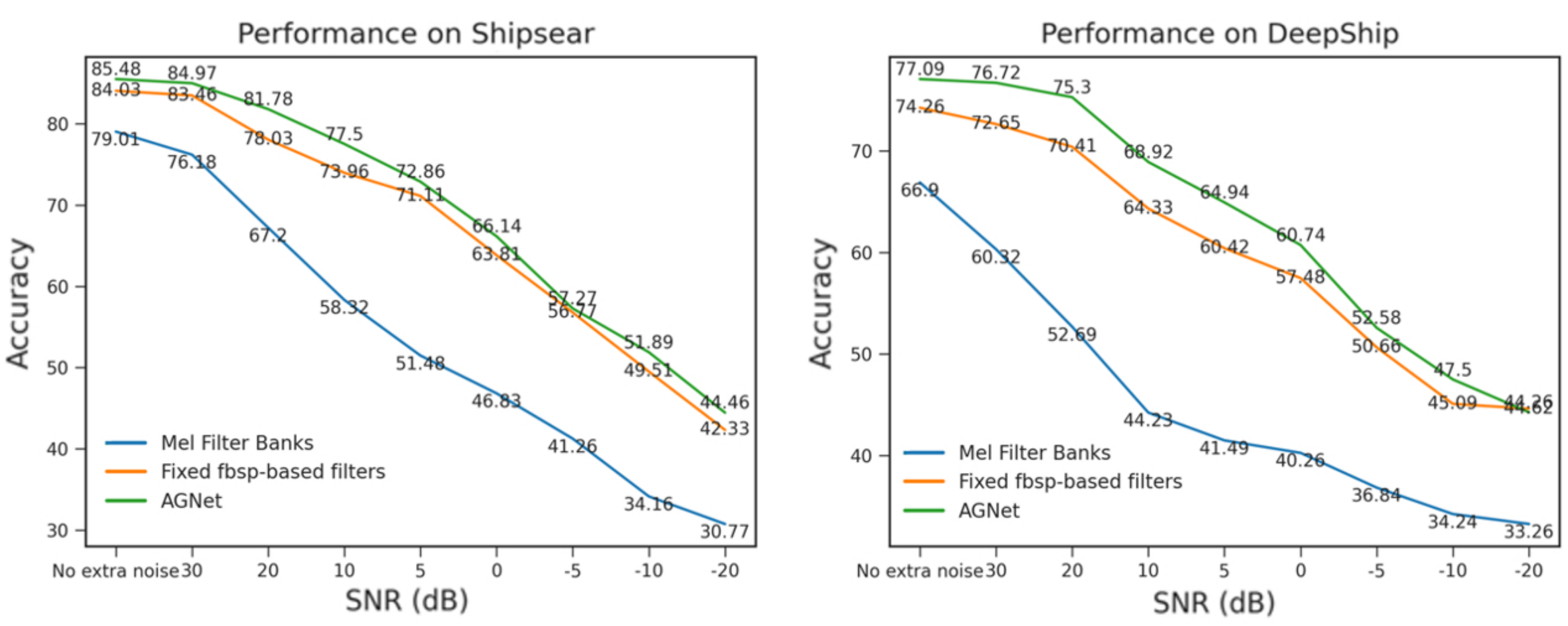}
        \centering
        \end{minipage}}
    }

    \scalebox{1}{
    \subfigure[Test on low cut-off frequency. On the left are the results on Shipsear and the right are the results on DeepShip.]{
        \begin{minipage}[b]{1\textwidth}
        \includegraphics[width=0.75\linewidth]{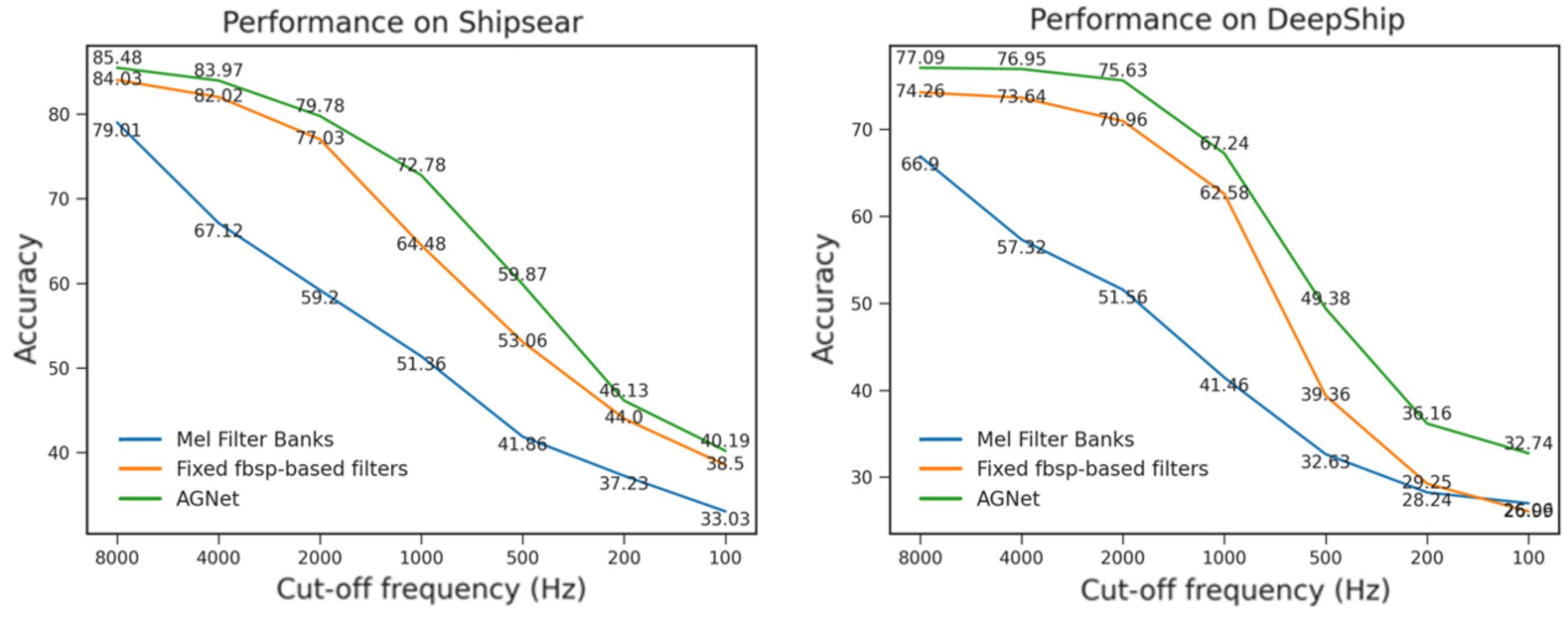}
        \centering
        \end{minipage}}
        }
    \caption{Robustness test on Signal to Noise Ratio and cut-off frequency.}
    \label{fig:robust}
\end{figure*}

Frequency truncation is performed on all Shipsear and DeepShip data. Figure~\ref{fig:robust}(b) indicates that the wavelet-based AGNet is appropriate for the scenarios that need to handle low cutoff frequency. For low-dimensional features such as Mel fbanks, the recognition accuracy could be approximately seen as a linear decrease with the decrease of the cut-off frequency, while the accuracy of the two wavelet spectrogram-based features decreases relatively slowly until the cut-off frequency reaches 2000 Hz. When the cut-off frequency decreases from 8000Hz to 2000Hz, AGNet has only 5.7\% and 1.46\% performance loss on Shipsear and DeepShip respectively. According to our analysis, with the decrease in cut-off frequency, the disadvantage of low-dimensional features with less information would be magnified. The lack of remaining information will bring challenges to the recognition system. For time-frequency-based features, the loss of a small part of the information has little effect on the neural network. Only when the remaining frequency band is too narrow (such as 0-500Hz) to capture enough information, the recognition ability of the network will drop sharply.

\section{Conclusion}

This work proposes an end-to-end ship-radiated noise recognition system - AGNet. By applying a fine-grained wavelet transform and a CNN-based classifier with parallel convolution attention blocks, AGNet achieves promising recognition accuracy on three datasets: Shipsear (85.48\%), DeepShip (77.09\%), and DTIL (95.76\%). This work innovatively introduces transfer learning to the underwater acoustic field. Moreover, experiments show that AGNet is robust against additive colored noise and low cut-off frequency. 

To conclude, AGNet could learn and update parameters in a data-driven manner. It could free researchers from tuning parameters, thus reducing troublesome work. Furthermore, such an end-to-end structure reduces the difficulty of actual deployment. It can save time consumption and reduces the possibility of module docking problems.

In the future, we plan to further explore the performance of adaptive learning on more complex data or tasks. For instance, setting more learnable parameters in the time-frequency feature extraction step.

\section*{Acknowledgements}
This research is partially supported by the IOA Frontier Exploration Project (No. ZYTS202001), the High Tech Project (No. 31513070501), Youth Innovation Promotion Association CAS.

\newpage

\printcredits

\bibliographystyle{cas-model2-names}

\bibliography{cas-refs-agnet}





\end{document}